\documentclass{iopart}
\usepackage{graphicx}
\begin{document}
\title{Global polarization measurement in Au+Au collisions}
\author{Ilya Selyuzhenkov for the STAR Collaboration}
\address{Physics Department, Wayne State University, 666 W Hancock, Detroit MI 48201, USA}
\ead{Ilya.Selyuzhenkov@wayne.edu}

\begin{abstract}
The system created in non central relativistic nucleus-nucleus collisions carries large angular orbital momentum.
Due to spin-orbital coupling, produced particles could be globally polarized along the direction of the system angular momentum. 
We present results of a measurement of Lambda hyperon global polarization in Au+Au collisions
at the center of mass energies 62 and 200 GeV with the STAR detector at RHIC.
The observed global polarization of Lambda
hyperons in the STAR acceptance is consistent with zero within the precision of the measurement.
The obtained upper limit,  $|P_{\Lambda}| < 0.01$,
is significantly below the theoretical values discussed recently in the literature.
\end{abstract}
\pacs{25.75.-q, 24.70.+s, 25.75.Ld}

\section{\label{Introduction}Introduction}

The system created in non central relativistic nucleus-nucleus collision carries large angular orbital momentum.
One of the most interesting and important phenomena predicted to occur in such a system is global system polarization.
The phenomena of global system polarization was first discussed in work \cite{LiangPRL94},
where theoretical calculation of the globally polarized system
produced in non central nucleus-nucleus collision was carried out.
The global polarization originates from transformation of the orbital angular momentum of the system into the particle's spin
due to spin-orbital coupling.
The latter leads to the polarization of secondary produced particles along the system orbital momentum
\cite{LiangPRL94, Voloshin0410089, Liang0411101}.
In this paper we present the results of $\Lambda$--hyperon global polarization measurement in Au+Au collisions
at $\sqrt{s_{NN}}$ = 62 and 200 GeV with the STAR detector at RHIC.

\section{\label{GlobalPolarization}Global polarization of hyperons}

Particles produced in a system with large angular orbital momentum {\boldmath $L$}
are predicted to be polarized along the 
normal to the collision reaction plane due to spin-orbital coupling \cite{LiangPRL94, Voloshin0410089, Liang0411101}.
This global polarization can be determined from the angular distribution of hyperons decay
product with respect to the system orbital momentum {\boldmath $L$}:
\begin{eqnarray}
\label{GlobalPolarizationDefinition}
\frac{dN}{d \cos\theta^*} \sim 1~+~\alpha_H~P_H~\cos \theta^* ~~,
\end{eqnarray}
where $P_H$ is the hyperon global polarization,
$\alpha_H$ is the hyperon's decay parameter,
$\theta^*$ is the angle between the system orbital momentum {\boldmath $L$}
and the hyperon's decay products 3-momentum in the hyperon's rest frame.

Since the source for the global polarization is the angular orbital momentum {\boldmath $L$}
and this angular momentum is perpendicular to the reaction plane,
the global polarization could be measured via correlations with respect to the reaction plane.
Thus the known and well established anisotropic flow measurement techniques \cite{Voloshin:1994mz, Poskanzer:1998yz} may be applied.

In order to write the equation for the global polarization in terms of the convenient observables used in anisotropic flow measurements
we start with an equation that directly follows from the global polarization definition (\ref{GlobalPolarizationDefinition}):
\begin{eqnarray}
\label{GlobalPolarizationMeanCos}
P_{H}~=~\frac{3}{\alpha_H}~\langle \cos \theta^*\rangle~~.
\end{eqnarray}
The brackets in this equation denotes averaging over all possible directions
of the hyperon's decay product 3-momentum in the hyperon's rest frame
and over all directions of the system orbital momentum {\boldmath $L$} or,
in other words, over all possible orientation of the reaction plane.
Similarly we can write the equation for the global polarization in terms of
reaction plane angle $\Psi_{RP}$ and the azimuthal angle $\phi^*_p$
of the hyperon's decay product 3-momentum in the hyperon's rest frame.
By using the geometrical relation between the angles, namely
$\cos \theta^* = \sin \theta^*_p \cdot \sin \left( \phi^*_p - \Psi_{RP}\right)$
($\theta^*_p$ is the angle between hyperon's decay product 3-momentum in the hyperon's rest frame and beam direction),
and integrating angular distribution (\ref{GlobalPolarizationDefinition}) over the angle $\theta^*_p$,
one ends up with the following equation for the global polarization:
\begin{eqnarray}
\label{GlobalPolarizationObservable}
P_{H}~=~\frac{8}{\pi\alpha_H}\langle \sin \left( \phi^*_p - \Psi_{RP}\right)\rangle~~.
\end{eqnarray}
Note that in this equation, perfect detector acceptance is assumed.
To take into account the hyperon's reconstruction acceptance one has to correct the result on
$(4/\pi)\int {d\Omega^*_p/(4\pi) \cdot \sin \theta^*_p}$,
where $d\Omega^*_p = d\phi^*_p \sin \theta^*_p d \theta^*_p$ and the integral extends over detector acceptance.
The equation (\ref{GlobalPolarizationObservable}) is very similar to those used in directed flow measurements
\cite{Barrette:1996rs, Alt:2003ab, Adams:2005aa, Adams:2005ca, BeltTonjes:2004jw}.
For example, the hyperon's directed flow $v_1^H$ could be measured as:
$v_1^H = \langle \cos \left( \phi_H - \Psi_{RP}\right)\rangle$, where $\phi_H$ is the angle of the hyperon's transverse momentum.
This similarity allows us to use anisotropic flow measurement techniques.
In this paper we follow the same naming conventions and notations adopted in anisotropic flow analysis.

\subsection{\label{Technique}Technique}

This analysis is based on $\Lambda$--hyperons reconstructed with the STAR main TPC.
The decay channel used to reconstruct $\Lambda$--hyperons in STAR is  $\Lambda \to p \pi^- $ \cite{Adler:2002pb, Cai:2005ph,Takahashi:2005pq};
the corresponding decay papameter is $\alpha_{\Lambda}^{-} = 0.642\pm0.013$ \cite{Eidelman:2004wy}.
The $\Lambda$ global polarization is calculated based on equation (\ref{GlobalPolarizationObservable}).
As in anisotropic flow measurements, one can {\it estimate} the reaction plane angle
based on the information on all produced particles by calculating the so called $n$-harmonic event plane vector $Q^n_{EP}$.
This implies the necessity to correct the final results by the event plane resolution $R^n_{EP}$ 
(see \cite{Voloshin:1994mz, Poskanzer:1998yz} for more details).
In the global polarization measurement, knowledge of the system orbital momentum {\boldmath $L$} orientation is required,
and thus it is required to obtain the first order event plane vector $Q^1_{EP}$.
Taking this into account one can write equation (\ref{GlobalPolarizationObservable})
for the case of $\Lambda$ global polarization in terms of the first order event plane angle $\Psi_{EP}^1$ and its resolution $R_{EP}^1$:
\begin{eqnarray}
\label{GlobalPolarizationScalarProduct}
P_{\Lambda}~=~\frac{8}{\pi\alpha_\Lambda^-}\frac{\langle \sin \left( \phi^*_p - \Psi_{EP}^1\right)\rangle}{R_{EP}^1}~~.
\end{eqnarray}
In this paper, the first order event plane vector is determined based on particles measured in STAR Forward TPC 
(covered pseudo-rapidity region is $2.7 < |\eta| < 3.9$).

\subsection{\label{UpperLimit}$\Lambda$-hyperon global polarization upper limit}

\begin{figure}[h]
\begin{center}
\includegraphics[width=0.7\textwidth]{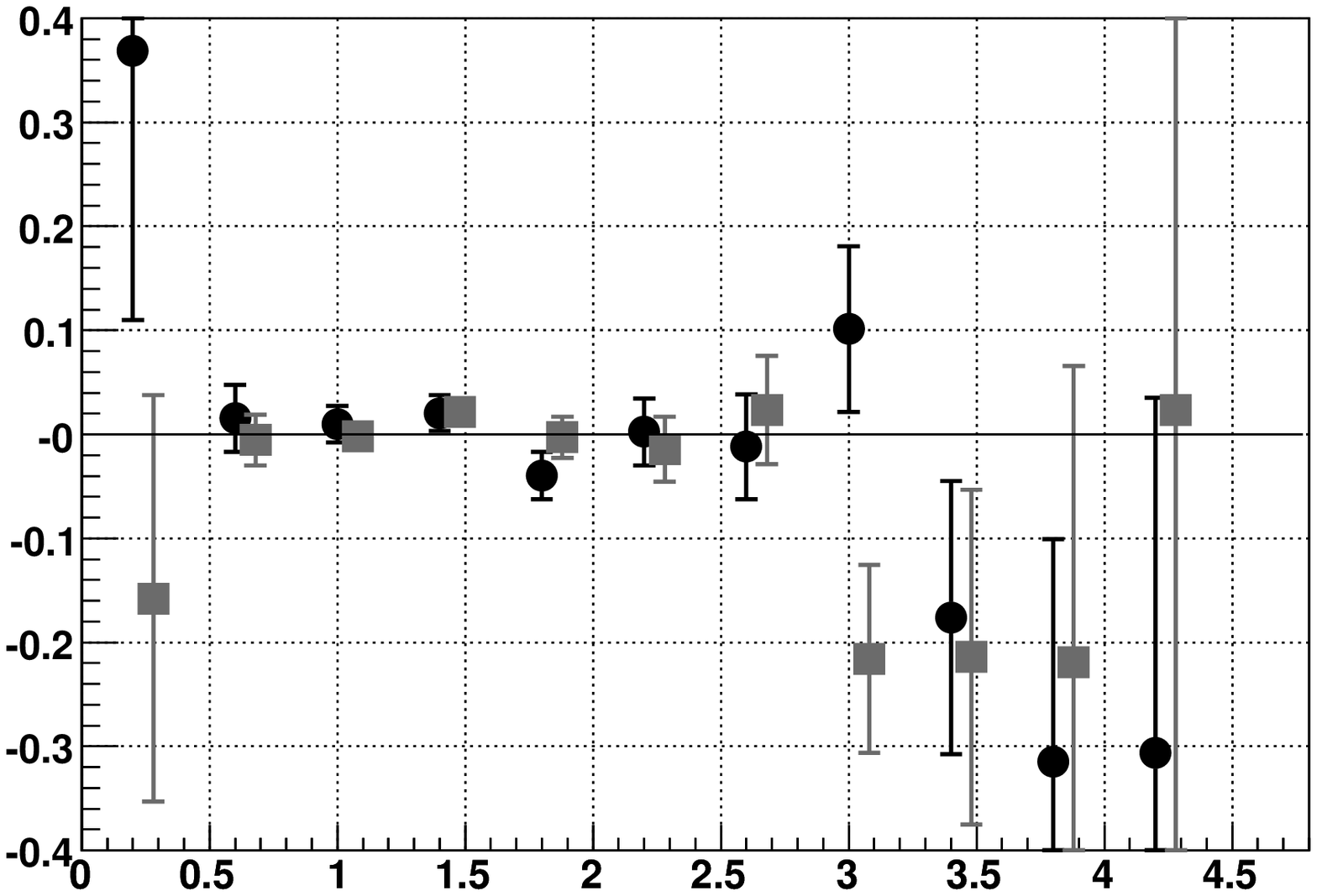}
\put(-265,82){\rotatebox{90}{$P_{\Lambda}$}}
\put(-160,145){{\bf STAR Preliminary}}
\put(-133,-5){$p_t^{\Lambda}$}
\caption{\label{lambdaGlobalPolarization_pt}
Global polarization of $\Lambda$--hyperons as a function of $\Lambda$ transverse momentum $p_t^\Lambda$.
Black circles show results from the measurement for Au+Au collisions at $\sqrt{s_{NN}}$=200 GeV (STAR centrality region 20-70\%).
Gray squares show results for Au+Au collisions at $\sqrt{s_{NN}}$=62 GeV (STAR centrality region 0-80\%).
Only statistical errors are shown.
}
\end{center}
\end{figure}

Fig.\ref{lambdaGlobalPolarization_pt} shows measured $\Lambda$--hyperon global polarization
as a function of $\Lambda$ transverse momentum $p_t^{\Lambda}$.
The black circles indicate results from the measurement based on the 8 million minimum bias events
obtained with the STAR detector during RHIC run IV (year 2004) for Au+Au collisions at $\sqrt{s_{NN}}$=200GeV.
The gray squares are from similar measurement based on the 9 million minimum bias events
for Au+Au collisions at $\sqrt{s_{NN}}$=62GeV.
Although the error bars at higher $\Lambda$ transverse momentum are rather large,
Fig.\ref{lambdaGlobalPolarization_pt} hints at a possible $p_t^\Lambda$ dependence of the global polarization.
Unfortunately, at present there is no available theory prediction of this dependence.

\begin{figure}[h]
\begin{center}
\includegraphics[width=0.7\textwidth]{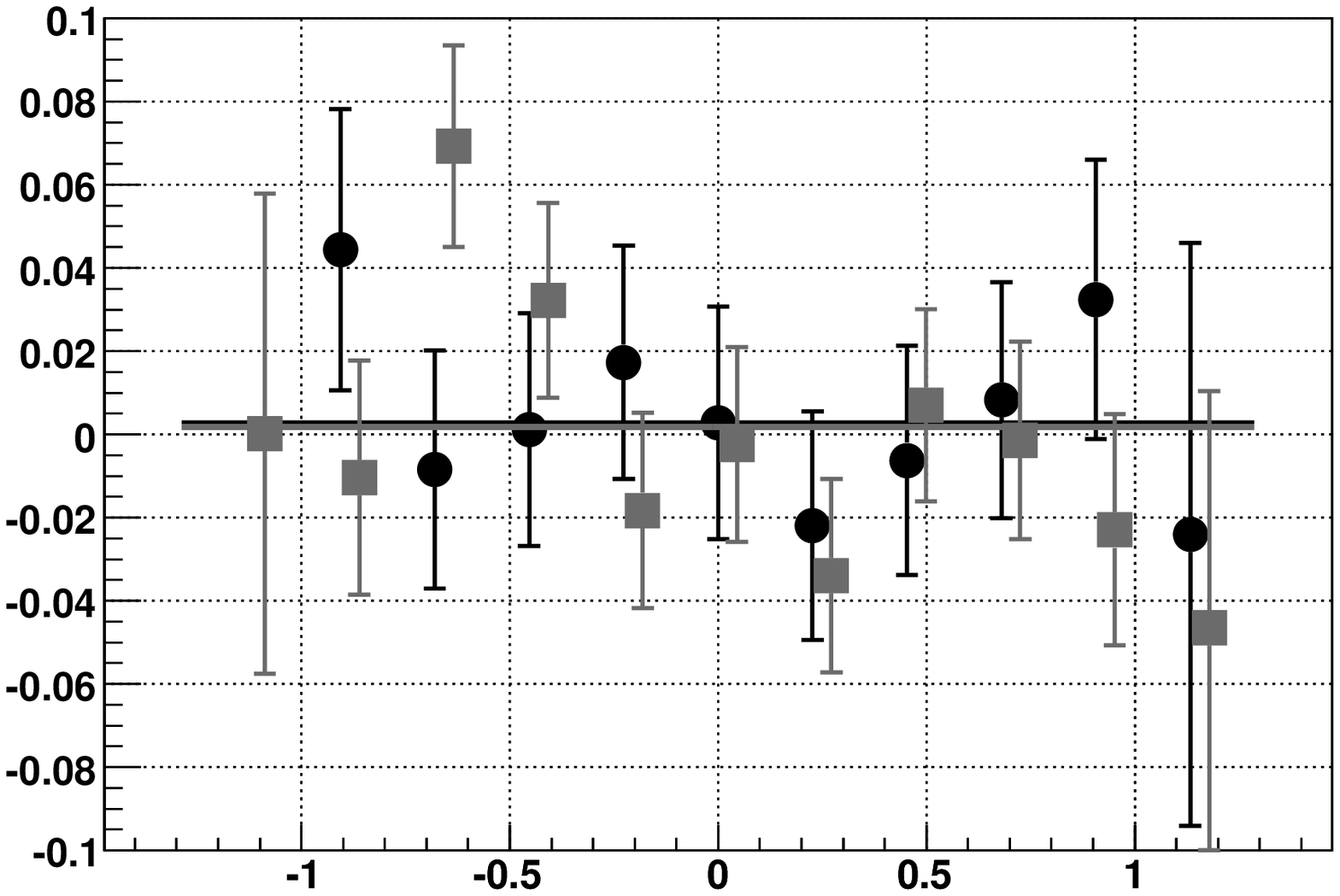}
\put(-265,82){\rotatebox{90}{$P_{\Lambda}$}}
\put(-160,145){{\bf STAR Preliminary}}
\put(-133,-5){$\eta^{\Lambda}$}
\caption{\label{lambdaGlobalPolarization_eta}
Global polarization of $\Lambda$--hyperons as a function of $\Lambda$ pseudo-rapidity $\eta^\Lambda$.
Black circles show results from the measurement for Au+Au collisions at $\sqrt{s_{NN}}$=200 GeV (STAR centrality region 20-70\%).
Constant line fit to this data points: $P_{\Lambda} = (2.6 \pm 9.5) \times 10^{-3}$.
Gray squares show results for Au+Au collisions at $\sqrt{s_{NN}}$=62 GeV (STAR centrality region 0-80\%).
Constant line fit gives: $P_{\Lambda} = (1.9 \pm 8.0) \times 10^{-3}$.
Only statistical errors are shown.
}
\end{center}
\end{figure}

Fig.\ref{lambdaGlobalPolarization_eta} presents the measured $\Lambda$--hyperon global polarization
as a function of $\Lambda$  pseudo-rapidity $\eta^{\Lambda}$.
Symbols are the same as in Fig.\ref{lambdaGlobalPolarization_pt}.
Note, that scale is different from the one in Fig.\ref{lambdaGlobalPolarization_pt}.
The $p_t$ integrated global polarization result
is dominated by the region $p^{\Lambda}_t<3$~GeV
where the $\Lambda$--hyperon global polarization is consistent with zero (see Fig.\ref{lambdaGlobalPolarization_pt}).
The solid lines correspond to fitting a line of zero slope to the experimental data
(black is the fit to the result for Au+Au collisions at $\sqrt{s_{NN}}$=200 GeV and
gray is the fit for the same collision system at $\sqrt{s_{NN}}$=62 GeV).
The results for the $\Lambda$--hyperon global polarization within
the STAR acceptance are consistent with zero.

\section{\label{Conclusion}Conclusion}

The $\Lambda$--hyperon global polarization has been measured in Au+Au collisions 
at the center of mass energies $\sqrt{s_{NN}}$=62 and 200 GeV with the STAR detector at RHIC.
An upper limit of $|P_{\Lambda}| < 0.01$ for the global polarization of $\Lambda$--hyperons within the STAR acceptance is found.
This value is far below the one discussed in the recent theoretical papers: $P_{\Lambda} = - 0.3$ \cite{LiangPRL94}.
The reason for this significant discrepancy is not clear now and there are still extensive theoretical discussions on this subject.
As it was found later by the original authors of \cite{LiangPRL94} the predicted value of $P_{\Lambda} = - 0.3$
could be incorrect due to inapplicability of the approximations used
and the correct estimation for RHIC energies requires more realistic theoretical calculations (see erratum in \cite{LiangPRL94} for details).

\end{document}